\documentclass[10pt,twocolumn,english,aps,preprint]{revtex4}
\usepackage[T1]{fontenc}
\usepackage[latin1]{inputenc}
\usepackage{graphicx}

\makeatletter

\usepackage{graphicx}

\usepackage{babel}
\makeatother
\begin{document}

\title{Group expansions for impurities in superconductors}

\author{Yu.G. Pogorelov}

\email{ypogorel@fc.up.ptv}

\affiliation{CFP, Universidade do Porto, Rua do Campo Alegre 687, 4169-007 Porto,
Portugal}

\author{V.M. Loktev}

\email{vloktev@bitp.kiev.ua}

\affiliation{Bogolyubov Institute for Theoretical Physics, 14b Metrologichna str.,
03143 Kiev, Ukraine}

\begin{abstract}
A new method is proposed for practical calculation of the effective
interaction between impurity scatterers in superconductors, based
on algebraic properties of related Nambu matrices for Green functions.
In particular, we show that the density of states within the s-wave
gap can have a non-zero contribution (impossible either in Born and
in T-matrix approximation) from non-magnetic impurities with concentration
$c\ll1$, beginning from $\sim c^{3}$ order. 
\end{abstract}
\maketitle

\section{introduction}

\label{sec:int}Impurity effects are at the center of interest in
studies of superconducting (SC) materials, especially of those with
high transition temperature (HTSC). In the general theory of disordered
systems with disorder due to diluted impurity centers, the so-called
group expansion (GE) method was proposed as most consistent for quasiparticle
Green function (GF) \cite{iv}, and it was also formulated for SC
systems \cite{pog}.

The GE's of different types (see below) are analogous to the classical
Ursell-Mayer group series in the theory of non-ideal gases \cite{may},
where the particular terms (the group integrals) include physical
interactions in groups of the given number of particles. In the quantum
theory of solids, GE includes indirect interactions (dependent on
the excitation energy $\varepsilon$) between the impurity centers,
through the exchange by virtual excitations from (admittedly renormalized)
band spectrum, so that each term corresponds to summation of certain
infinite series of diagrams. These expansions were elaborated in a
detail for various kinds of normal quasiparticle spectra, where they
define the interplay between extended and localized states \cite{ilp},
however their usage in SC systems encounters considerable technical
difficulties due to existence of anomalous GF's.

The present paper is aimed on an efficient algorithm for resolving
these difficulties. We develop the specific algebraic techniques to
calculate Nambu matrices in various terms of GE for the exemplary
case of s-wave symmetry of SC order parameter, permitting to explore
the impurity effects in this case beyond the scope of Anderson theorem
\cite{and}. In particular, we find that pair clusters of impurities
(2nd term of GE) can not produce finite contribution into the quasiparticle
density of states (DOS) within the s-wave gap, alike the simplest
effect of isolated impurities (1st GE term), but a non-zero contribution
into the in-gap DOS is already possible for the 3rd GE term (impurity
triples). These results allow also a straightforward extension to
the d-wave symmetry characteristic for doped HTSC materials (where
the dopants not only supply the charge carriers but also play the
role of impurity scatterers).

\section{Hamiltonian and Green functions}

\label{sec:ham}For description of electronic spectra in a SC system
with impurities, it is convenient to use the formalism of Nambu spinors:
the row-spinor $\psi_{\mathbf{k}}^{\dagger}=(a_{\mathbf{k},\uparrow}^{\dagger},a_{-\mathbf{k},\downarrow})$
and respective column-spinor $\psi_{\mathbf{k}}$, writing the Hamiltonian
in a spinor form \begin{equation}
H_{sc}=\sum_{\mathbf{k}}[\psi_{\mathbf{k}}^{\dagger}(\xi_{\mathbf{k}}\widehat{\tau}_{3}+\Delta_{\mathbf{k}}\widehat{\tau}_{1})\psi_{\mathbf{k}}-\frac{1}{N}\sum_{\mathbf{p},\mathbf{k}^{\prime}}\mathrm{e}^{i(\mathbf{k}-\mathbf{k}^{\prime})\mathbf{p}}\psi_{\mathbf{k}^{\prime}}^{\dagger}\widehat{V}\psi_{\mathbf{k}}].\label{eq:1}\end{equation}
 It includes the normal quasiparticle energy $\xi_{\mathbf{k}}$,
the mean-field gap function $\Delta_{\mathbf{k}}$, the Pauli matrices
$\widehat{\tau}_{j}$ and the perturbation matrix $\widehat{V}=V_{\mathrm{L}}\widehat{\tau}_{3}$.
The impurity (attractive) perturbation on random sites $\mathbf{p}$
with concentration $c=\sum_{\mathbf{p}}1/N\ll1$ is described by the
Lifshitz parameter $V_{\mathrm{L}}$.

The energy spectrum of a Fermi system is described through the Fourier
transformed two-time Green functions (GF's) \cite{eco}: \begin{equation}
\langle\langle a|b\rangle\rangle_{\varepsilon}=i\int_{-\infty}^{0}{\textrm{e}}^{i(\varepsilon-i0)t}\langle\{ a(t),b(0)\}\rangle dt,\label{eq:2}\end{equation}
 where $\langle\ldots\rangle$ is the quantum statistical average
and $\{.,.\}$ is the anticommutator of operators in Heisenberg representation.
For the system, Eq. \ref{eq:1}, we define the $2\times2$ Nambu matrix
of GF's \begin{equation}
\widehat{G}_{\mathbf{k},\mathbf{k}^{\prime}}=\langle\langle\psi_{\mathbf{k}}|\psi_{\mathbf{k}^{\prime}}^{\dagger}\rangle\rangle.\label{eq:3}\end{equation}
 The matrix elements in the expanded form of Eq. \ref{eq:3} are the
well-known Gor'kov normal and anomalous functions \cite{gor}. In
what follows, we shall also distinguish between the Nambu indices
(N-indices) and the quasi-momentum indices (m-indices) in this matrix.
In absence of impurities, the explicit form of the matrix, Eq. \ref{eq:3},
turns: $\widehat{G}_{\mathbf{k},\mathbf{k}^{\prime}}\rightarrow\delta_{\mathbf{k},\mathbf{k}^{\prime}}\widehat{G}_{\mathbf{k}}^{0}$,
where the non-perturbed (m-diagonal) GF matrix \begin{equation}
\widehat{G}_{\mathbf{k}}^{0}=\frac{\varepsilon+\xi_{\mathbf{k}}\widehat{\tau}_{3}+\Delta_{\mathbf{k}}\widehat{\tau}_{1}}{\varepsilon^{2}-E_{\mathbf{k}}^{2}}\label{eq:4}\end{equation}
 involves the SC quasiparticle energy $E_{\mathbf{k}}=\sqrt{\xi_{\mathbf{k}}^{2}+\Delta_{\mathbf{k}}^{2}}$.
The relevant physical properties of SC state are suitably expressed
in terms of these GF's. For instance, the single-particle DOS, related
to the electronic specific heat, is given by \begin{equation}
\rho(\varepsilon)=\frac{1}{\pi N}\sum_{\mathbf{k}}\mathrm{Tr}\:{\textrm{Im}}\widehat{G}_{\mathbf{k}}\label{eq:5}\end{equation}
 where $\widehat{G}_{\mathbf{k}}\equiv\widehat{G}_{\mathbf{k},\mathbf{k}}$
is the m-diagonal GF.

Now we pass to calculation of GF's in SC systems at finite concentration
$c$ of impurity centers and analyze explicit structure of corresponding
GE's.

\section{Group expansions for self-energy}

\label{sec:ge}We derive GE's for the system defined by the Hamiltonian
Eq. \ref{eq:1}, starting from the Dyson equation of motion for a
matrix GF: \begin{equation}
\widehat{G}_{\mathbf{k},\mathbf{k}^{\prime}}=\widehat{G}_{\mathbf{k}}^{0}\delta_{\mathbf{k},\mathbf{k}^{\prime}}-\frac{1}{N}\sum_{\mathbf{p},\mathbf{k}^{\prime\prime}}\mathrm{e}^{i(\mathbf{\mathbf{k}^{\prime\prime}}-\mathbf{k}^{\prime})\mathbf{p}}\widehat{G}_{\mathbf{k}}^{0}\widehat{V}\widehat{G}_{\mathbf{k}^{\prime\prime},\mathbf{k}^{\prime}}.\label{eq:6}\end{equation}
 A routine consists in consecutive iterations of the same equations
for the GF's in the {}``scattering'' terms of Eq. \ref{eq:6} and
separating systematically those already present in the previous iterations
\cite{iv}. Thus, for the m-diagonal GF $\widehat{G}_{\mathbf{k}}$,
we first separate the scattering term with the function $\widehat{G}_{\mathbf{k}}$
itself from those with $\widehat{G}_{\mathbf{k}^{\prime},\mathbf{k}}$,
$\mathbf{k}^{\prime}\neq\mathbf{k}$: \begin{eqnarray}
 &  & \widehat{G}_{\mathbf{k}}=\widehat{G}_{\mathbf{k}}^{0}-\frac{1}{N}\sum_{\mathbf{k}^{\prime},\mathbf{p}}\mathrm{e}^{i\left(\mathbf{k}-\mathbf{k}^{\prime}\right)\cdot\mathbf{p}}\widehat{G}_{\mathbf{k}}^{0}\widehat{V}\widehat{G}_{\mathbf{k}^{\prime},\mathbf{k}}=\nonumber \\
 &  & =\widehat{G}_{\mathbf{k}}^{0}-c\widehat{G}_{\mathbf{k}}^{0}\widehat{V}\widehat{G}_{\mathbf{k}}-\frac{1}{N}\sum_{\mathbf{k}^{\prime}\neq\mathbf{k},\mathbf{p}}\mathrm{e}^{i\left(\mathbf{k}-\mathbf{k}^{\prime}\right)\cdot\mathbf{p}}\widehat{G}_{\mathbf{k}}^{0}\widehat{V}\widehat{G}_{\mathbf{k}^{\prime},\mathbf{k}}.\label{eq:7}\end{eqnarray}
 Then for each $\widehat{G}_{\mathbf{k}^{\prime},\mathbf{k}}$, $\mathbf{k}^{\prime}\neq\mathbf{k}$
we write down Eq. \ref{eq:6} again and single out the scattering
terms with $\widehat{G}_{\mathbf{k}}$ and $\widehat{G}_{\mathbf{k}^{\prime},\mathbf{k}}$
in its r.h.s: \begin{eqnarray}
\widehat{G}_{\mathbf{k}^{\prime},\mathbf{k}}=-\frac{1}{N}\sum_{\mathbf{k}^{\prime\prime},\mathbf{p}^{\prime}}\mathrm{e}^{i\left(\mathbf{k}^{\prime}-\mathbf{k}^{\prime\prime}\right)\cdot\mathbf{p}}\widehat{G}_{\mathbf{k}^{\prime}}^{0}\widehat{V}\widehat{G}_{\mathbf{k}^{\prime\prime},\mathbf{k}}=\nonumber \\
=-c\widehat{G}_{\mathbf{k}^{\prime}}^{0}\widehat{V}\widehat{G}_{\mathbf{k}^{\prime},\mathbf{k}}-\frac{1}{N}\mathrm{e}^{i\left(\mathbf{k}^{\prime}-\mathbf{k}\right)\cdot\mathbf{p}}\widehat{G}_{\mathbf{k}^{\prime}}^{0}\widehat{V}\widehat{G}_{\mathbf{k}}\nonumber \\
-\frac{1}{N}\sum_{\mathbf{p^{\prime}\neq p}}\mathrm{e}^{i\left(\mathbf{k}^{\prime}-\mathbf{k}\right)\cdot\mathbf{p}^{\prime}}\widehat{G}_{\mathbf{k}^{\prime}}^{0}\widehat{V}\widehat{G}_{\mathbf{k}}\nonumber \\
-\frac{1}{N}\sum_{\mathbf{k}^{\prime\prime}\neq\mathbf{k},\mathbf{k}^{\prime};\mathbf{p}^{\prime}}\mathrm{e}^{i\left(\mathbf{k}^{\prime}-\mathbf{k}^{\prime\prime}\right)\cdot\mathbf{p}^{\prime}}\widehat{G}_{\mathbf{k}^{\prime}}^{0}\widehat{V}\widehat{G}_{\mathbf{k}^{\prime\prime},\mathbf{k}}.\label{eq:8}\end{eqnarray}
 Note that among the terms with $\widehat{G}_{\mathbf{k}}$, the $\mathbf{p}^{\prime}=\mathbf{p}$
term (the second in r.h.s. of Eq. \ref{eq:8}) bears the phase factor
$\mathrm{e}^{i\left(\mathbf{k}^{\prime}-\mathbf{k}\right)\cdot\mathbf{p}}$,
so it is coherent to that already figured in the last sum in Eq. \ref{eq:7}.
That is why this term is explicitly separated from other, incoherent
ones, $\mathrm{\propto e}^{i\left(\mathbf{k}^{\prime}-\mathbf{k}\right)\cdot\mathbf{p}^{\prime}}$,
$\mathbf{p}^{\prime}\neq\mathbf{p}$ (but there will be no such separation
when doing 1st iteration of Eq. \ref{eq:6} for the m-non-diagonal
GF $\widehat{G}_{\mathbf{k}^{\prime\prime},\mathbf{k}}$ itself).

Continuing the sequence, we collect the terms with the initial function
$\widehat{G}_{\mathbf{k}}$which result from: i) all multiple scatterings
on the same site $\mathbf{p}$ and ii) such processes on the same
pair of sites $\mathbf{p}$ and $\mathbf{p}^{\prime}\neq\mathbf{p}$.
Then summation in $\mathbf{p}$ of the i)-terms gives rise to the
first term of GE, and, if the pair processes were neglected, it would
coincide with the well known result of self-consistent T-matrix approximation
\cite{baym}. The second term of GE, obtained by summation in $\mathbf{p},\mathbf{p}^{\prime}\neq\mathbf{p}$
of the ii)-terms, contains certain interaction matrices $\widehat{A}_{\mathbf{p}^{\prime},\mathbf{p}}$
generated by the multiply scattered functions $\widehat{G}_{\mathbf{k}^{\prime},\mathbf{k}}$,
$\mathbf{k}^{\prime}\neq\mathbf{k}$, etc., (including their own renormalization).
For instance, the iterated equation of motion for a function $\widehat{G}_{\mathbf{k}^{\prime\prime},\mathbf{k}}$
with $\mathbf{k}^{\prime\prime}\neq\mathbf{k},\mathbf{k}^{\prime}$
in the last term of Eq. \ref{eq:7} will produce: \begin{eqnarray}
 &  & \widehat{G}_{\mathbf{k}^{\prime\prime},\mathbf{k}}=-\frac{1}{N}\sum_{\mathbf{k}^{\prime\prime\prime},\mathbf{p}^{\prime\prime}}\mathrm{e}^{i\left(\mathbf{k}^{\prime\prime}-\mathbf{k}^{\prime\prime\prime}\right)\cdot\mathbf{p}^{\prime\prime}}\widehat{G}_{\mathbf{k}^{\prime\prime}}^{0}\widehat{V}\widehat{G}_{\mathbf{k}^{\prime\prime\prime},\mathbf{k}}=\nonumber \\
 &  & =-\frac{\mathrm{e}^{i\left(\mathbf{k}^{\prime\prime}-\mathbf{k}\right)\cdot\mathbf{p}}}{N}\widehat{G}_{\mathbf{k}^{\prime\prime}}^{0}\widehat{V}\widehat{G}_{\mathbf{k}}-\frac{\mathrm{e}^{i\left(\mathbf{k}^{\prime\prime}-\mathbf{k}\right)\cdot\mathbf{p}^{\prime}}}{N}\widehat{G}_{\mathbf{k}^{\prime\prime}}^{0}\widehat{V}\widehat{G}_{\mathbf{k}}+\nonumber \\
 &  & +\quad\mathrm{terms\quad with}\quad\widehat{G}_{\mathbf{k}^{\prime},\mathbf{k}}\quad\mathrm{and}\quad\widehat{G}_{\mathbf{k}^{\prime\prime},\mathbf{k}}+\nonumber \\
 &  & +\quad\mathrm{terms\quad with}\quad\widehat{G}_{\mathbf{k}^{\prime\prime\prime},\mathbf{k}}\;(\mathbf{k}^{\prime\prime\prime}\neq\mathbf{k},\mathbf{k}^{\prime},\mathbf{k}^{\prime\prime}).\label{eq:9}\end{eqnarray}
 Consequently, we obtain the solution for an m-diagonal GF as \begin{equation}
\widehat{G}_{\mathbf{k}}=\widehat{G}_{\mathbf{k},\mathbf{k}}=\left[\left(\widehat{G}_{\mathbf{k}}^{0}\right)^{-1}-\widehat{\Sigma}_{\mathbf{k}}\right]^{-1},\label{eq:10}\end{equation}
 with the matrix GE for the renormalized self-energy matrix: \begin{eqnarray}
 &  & \widehat{\Sigma}_{\mathbf{k}}=c\widehat{T}\left[1+c\sum_{\mathbf{n}\neq0}\left(\widehat{A}_{0,\mathbf{n}}{\textrm{e}}^{-i\mathbf{kn}}+\widehat{A}_{0,\mathbf{n}}\widehat{A}_{\mathbf{n},0}\right)\times\right.\nonumber \\
 &  & \left.\times\left(1-\widehat{A}_{0,\mathbf{n}}\widehat{A}_{\mathbf{n},0}\right)^{-1}+\cdots\right].\label{eq:11}\end{eqnarray}
 Here $\widehat{T}=-\widehat{V}\left(1+\widehat{G}\widehat{V}\right)^{-1}$
is the renormalized T-matrix, and the indirect interaction (mediated
by the quasiparticles of host crystal) between two impurities at lattice
sites $0$ and $\mathbf{n}$ is described by the matrix $\widehat{A}_{0,\mathbf{n}}=N^{-1}\sum_{\mathbf{k}^{\prime}\neq\mathbf{k}}{\textrm{e}}^{i\mathbf{\mathbf{k}^{\prime}n}}\widehat{G}_{\mathbf{k}^{\prime}}\widehat{T}$,
with the sum in quasi-momenta restricted due to the above algorithm
of separation. There are even more such restrictions in each product
of these matrices: $\widehat{A}_{0,\mathbf{n}}\widehat{A}_{\mathbf{n},0}=N^{-2}\sum_{\mathbf{k}^{\prime}\neq\mathbf{k}}\sum_{\mathbf{k}^{\prime\prime}\neq\mathbf{k},\mathbf{k}^{\prime}}{\textrm{e}}^{i\mathbf{\left(\mathbf{k}^{\prime}-\mathbf{k}^{\prime\prime}\right)n}}\widehat{G}_{\mathbf{k}^{\prime}}\widehat{T}\widehat{G}_{\mathbf{k}^{\prime\prime}}\widehat{T}$,
and so on. This seems to seriously hamper calculation of the sum $\sum_{\mathbf{n}\neq0}$
in Eq. \ref{eq:11} (not to say about higher GE terms). However, the
difficulty is avoided, taking into account the identities for two
first terms of its expansion \cite{iv}: \begin{eqnarray*}
\sum_{\mathbf{n}\neq0}\widehat{A}_{0,\mathbf{n}}{\textrm{e}}^{-i\mathbf{kn}}=-\widehat{A}_{0,0}+\sum_{\mathbf{n}}\widehat{A}_{0,\mathbf{n}}{\textrm{e}}^{-i\mathbf{kn}}=\\
=-\widehat{A}_{0,0}+\frac{1}{N}\sum_{\mathbf{n}}\sum_{\mathbf{k}^{\prime}\neq\mathbf{k}}{\textrm{e}}^{i\mathbf{\left(\mathbf{k}^{\prime}-\mathbf{k}\right)n}}\widehat{G}_{\mathbf{k^{\prime}}}\widehat{T}=\\
=-\widehat{A}_{0,0},\end{eqnarray*}
 and \begin{eqnarray*}
\sum_{\mathbf{n}\neq0}\widehat{A}_{0,\mathbf{n}}\widehat{A}_{\mathbf{n},0}=-\widehat{A}_{0,0}^{2}+\sum_{\mathbf{n}}\widehat{A}_{0,\mathbf{n}}\widehat{A}_{\mathbf{n},0}=-\widehat{A}_{0,0}^{2}+\\
+\frac{1}{N^{2}}\sum_{\mathbf{n}}\sum_{\mathbf{k}^{\prime},\mathbf{k}^{\prime\prime}\neq\mathbf{k}^{\prime}}{\textrm{e}}^{i\mathbf{\left(\mathbf{k}^{\prime}-\mathbf{k}^{\prime\prime}\right)n}}\widehat{G}_{\mathbf{k^{\prime}}}\widehat{T}\widehat{G}_{\mathbf{k}^{\prime\prime}}\widehat{T}=\\
=-\widehat{A}_{0,0}^{2},\end{eqnarray*}
 due to the momentum independence of T-matrix, and the fact that the
restrictions can be simply ignored in the higher products, like \begin{eqnarray*}
 &  & \sum_{\mathbf{n}\neq0}\widehat{A}_{0,\mathbf{n}}\widehat{A}_{\mathbf{n},0}\widehat{A}_{0,\mathbf{n}}{\textrm{e}}^{-i\mathbf{kn}}=-\widehat{A}_{0,0}^{3}+\\
 &  & +\sum_{\mathbf{n}}\widehat{A}_{0,\mathbf{n}}\widehat{A}_{\mathbf{n},0}\widehat{A}_{0,\mathbf{n}}{\textrm{e}}^{-i\mathbf{kn}}=-\widehat{A}_{0,0}^{3}+\\
 &  & \frac{1}{N^{3}}\sum_{\mathbf{n}}\sum_{\mathbf{k}^{\prime},\mathbf{k}^{\prime\prime}\neq\mathbf{k}^{\prime}}{\textrm{e}}^{i\mathbf{\left(\mathbf{k}^{\prime}-\mathbf{k}^{\prime\prime}+\mathbf{k}^{\prime\prime\prime}-\mathbf{k}\right)n}}\widehat{G}_{\mathbf{k^{\prime}}}\widehat{T}\widehat{G}_{\mathbf{k}^{\prime\prime}}\widehat{T}\widehat{G}_{\mathbf{k}^{\prime\prime\prime}}\widehat{T}=\\
 &  & -\widehat{A}_{0,0}^{3}+\frac{1}{N^{2}}\sum_{\mathbf{k}^{\prime},\mathbf{k}^{\prime\prime}}\widehat{G}_{\mathbf{k^{\prime}}}\widehat{T}\widehat{G}_{\mathbf{k}^{\prime\prime}}\widehat{T}\widehat{G}_{\mathbf{k}-\mathbf{k^{\prime}}+\mathbf{k}^{\prime\prime}}\widehat{T},\end{eqnarray*}
 etc. Thus we arrive at the final form for the renormalized GE \begin{eqnarray}
 &  & \widehat{\Sigma}_{\mathbf{k}}=c\widehat{T}\left[1-c\widehat{A}_{0,0}-c\widehat{A}_{0,0}^{2}+\right.\nonumber \\
 &  & \left.+c\sum_{\mathbf{n}\neq0}\left(\widehat{A}_{0,\mathbf{n}}^{3}{\textrm{e}}^{-i\mathbf{kn}}+\widehat{A}_{0,\mathbf{n}}^{4}\right)\left(1-\widehat{A}_{0,\mathbf{n}}^{2}\right)^{-1}+\cdots\right]\label{eq:11a}\end{eqnarray}
 where $\widehat{A}_{0,\mathbf{n}}=\widehat{G}_{0,\mathbf{n}}\widehat{T}$
and the renormalized local GF matrices $\widehat{G}_{0,\mathbf{n}}=N^{-1}\sum_{\mathbf{k}}{\textrm{e}}^{i\mathbf{kn}}\widehat{G}_{\mathbf{k}}$
and $\widehat{G}=\widehat{G}_{0,0}$ are already free from restrictions.
The two terms, next to unity in the brackets in Eq. \ref{eq:11a},
correspond to the excluded double occupancy of the same site by impurities,
the sum in $\mathbf{n}\neq0$ describes the averaged contribution
of all possible impurity pairs, and the dropped terms are for triples
and more of impurities.

An alternative routine for Eq. \ref{eq:6} consists in its iteration
for \emph{all} the terms $\widehat{G}_{\mathbf{k}^{\prime\prime},\mathbf{k}}$
and summing the contributions $\propto\widehat{G}_{\mathbf{k}}^{0}$,
like the first term in r.h.s. of Eq. \ref{eq:6}. This finally leads
to the solution of form \begin{equation}
\widehat{G}_{\mathbf{k}}=\widehat{G}_{\mathbf{k}}^{0}+\widehat{G}_{\mathbf{k}}^{0}\widehat{\Sigma}_{\mathbf{k}}^{0}\widehat{G}_{\mathbf{k}}^{0},\label{eq:12}\end{equation}
 with the non-renormalized self-energy matrix \begin{eqnarray}
 &  & \widehat{\Sigma}_{\mathbf{k}}^{0}=c\widehat{T}^{0}\left\{ 1+c\sum_{\mathbf{n}\neq0}\left[\widehat{A}_{0,\mathbf{n}}^{0}{\textrm{e}}^{-i\mathbf{kn}}+\left(\widehat{A}_{0,\mathbf{n}}^{0}\right)^{2}\right]\times\right.\nonumber \\
 &  & \left.\times\left[1-\left(\widehat{A}_{0,\mathbf{n}}^{0}\right)^{2}\right]^{-1}+\cdots\right\} ,\label{eq:13}\end{eqnarray}
 and the respective elements $\widehat{T}^{0}=-\widehat{V}\left(1+\widehat{G}^{0}\widehat{V}\right)^{-1}$,
$\widehat{A}_{0,\mathbf{n}}^{0}=\widehat{G}_{0,\mathbf{n}}^{0}\widehat{T}^{0}$,
$\widehat{G}_{0,\mathbf{n}}^{0}=N^{-1}\sum_{\mathbf{k}}{\textrm{e}}^{i\mathbf{kn}}\widehat{G}_{\mathbf{k}}^{0}$
and $\widehat{G}^{0}=\widehat{G}_{0,0}^{0}$.

Presenting GF's in the disordered system in the form of GE's generally
leads to respective expansions for its observable characteristics.
For instance, the impurity perturbed DOS is expected in the form:
$\rho\left(\varepsilon\right)=\rho_{0}\left(\varepsilon\right)+\rho_{1}\left(\varepsilon\right)+\rho_{2}\left(\varepsilon\right)+\ldots$,
related to contributions of pure crystal, isolated impurities, impurity
pairs, etc.

However, usage of each type of GE, the renormalized Eq. \ref{eq:11a}
or the non-renormalized Eq. \ref{eq:13}, is only justified if they
are convergent (at least, asymptotically). Since the matrices $\widehat{T}$
and $\widehat{A}$ are energy dependent, convergence of each type
of GE is restricted to certain energy ranges, and these ranges are
generally different. For a number of normal systems with impurities,
where GE's are constructed of scalar functions $A_{0,\mathbf{n}}$,
it was shown that the renormalized GE converges within the region
of band-like states, well characterized by the wave-vector, and the
non-renormalized GE does within the region of localized states \cite{ilp}.
To get quantitative estimates of convergence and higher order contributions
to self-energy, operating with the matrix functions $\widehat{A}_{0,\mathbf{n}}$
in Eqs. \ref{eq:11a} and \ref{eq:13}, a special techniques is necessary
that we construct below.

\section{Algebraic techniques for Nambu matrices}

\label{sec:imp3}Let us explicitly calculate the elements of above
defined GE's for the simplest s-wave symmetry of the SC gap function:
$\Delta_{s}\left(\mathbf{k}\right)=\Delta$. The unperturbed local
GF matrix is obtained as an expansion in Pauli matrices: \begin{equation}
\widehat{G}^{0}=-\left(\varepsilon+\Delta\widehat{\tau}_{1}\right)g_{0}-g_{3}\widehat{\tau}_{3},\label{eq:14}\end{equation}
 where \[
g_{0}\left(\varepsilon\right)=\frac{\pi\rho_{{\textrm{F}}}}{2\sqrt{\Delta^{2}-\varepsilon^{2}}},\]
 defines the s-wave DOS in pure crystal: $\rho_{0}\left(\varepsilon\right)=(2/\pi){\textrm{Im}}g_{0}\left(\varepsilon\right)$
(with the Fermi DOS $\rho_{{\textrm{F}}}$ of normal quasiparticles),
and the electron-hole asymmetry factor\[
g_{3}\left(\varepsilon\right)=\frac{1}{N}\sum_{\mathbf{k}}\frac{\xi_{\mathbf{k}}}{E_{\mathbf{k}}^{2}-\varepsilon^{2}}\approx\frac{1}{N}\sum_{\mathbf{k}}\frac{\xi_{\mathbf{k}}}{E_{\mathbf{k}}^{2}}\sim\rho_{{\textrm{F}}},\]
 is almost constant and real.

Then we readily calculate the non-renormalized T-matrix:\begin{equation}
\widehat{T}^{0}=\frac{2}{\pi\rho_{{\textrm{F}}}}\frac{v}{1+v^{2}}\left(\widehat{\tau}_{3}+v\frac{\varepsilon-\Delta\widehat{\tau}_{1}}{\sqrt{\Delta^{2}-\varepsilon^{2}}}\right),\label{eq:15}\end{equation}
 with the dimensionless perturbation parameter\[
v=\frac{\pi}{2}\frac{V_{\mathrm{L}}\rho_{{\textrm{F}}}}{1-V_{\mathrm{L}}g_{3}}.\]
 Since the denominator $1+v^{2}$ in Eq. \ref{eq:15} can not be zero,
the quasiparticle localization on a single impurity center in the
considered s-wave superconductor turns impossible \cite{pog}. If
the self-energy is approximated by the first term of GE: $\widehat{\Sigma}_{\mathbf{k}}\approx c\widehat{T}^{0}$,
then Eq. \ref{eq:15} used in Eq. \ref{eq:12} and in Eq. \ref{eq:5}
leads to the same DOS $\rho_{0}\left(\varepsilon\right)$ (that is,
$\rho_{1}\left(\varepsilon\right)\equiv0$) with the same gap value
$\Delta$ as in pure crystal. This justifies Anderson's theorem \cite{and}
within T-matrix approximation for an s-wave SC with point-like impurity
perturbations.

However, even if there is no in-gap poles in the single-impurity T-matrix
term, they can appear in the following terms of GE, that would describe
localized states on impurity clusters \cite{ilp}. Thus, the pair
contribution: \begin{eqnarray}
 &  & \rho_{2}\left(\varepsilon\right)=\frac{c^{2}}{\pi N}\sum_{\mathbf{k},\mathbf{n}\neq0}\mathrm{Tr\: Im}\widehat{G}_{\mathbf{k}}^{0}\widehat{T}^{0}\left[\widehat{A}_{0,\mathbf{n}}^{0}\cos\mathbf{kn}+\left(\widehat{A}_{0,\mathbf{n}}^{0}\right)^{2}\right]\times\nonumber \\
 &  & \times\left[1-\left(\widehat{A}_{0,\mathbf{n}}^{0}\right)^{2}\right]^{-1}\widehat{G}_{\mathbf{k}}^{0}\widehat{\tau}_{3},\label{eq:16}\end{eqnarray}
 should only follow from the poles of $[1-(\widehat{A}_{0,\mathbf{n}}^{0})^{2}]^{-1}$,
since the matrix $\widehat{A}_{0,\mathbf{n}}^{0}\left(\varepsilon\right)$
is real at $\varepsilon^{2}<$ $\Delta^{2}$.

In fact, the long distance asymptotics of this matrix (at $n\gg a$)
is:\begin{equation}
\widehat{A}_{0,\mathbf{n}}^{0}\approx-\mathcal{F}_{0,\mathbf{n}}\left(\varepsilon\right)\left(\cos\varphi_{\mathbf{n}}+\frac{\varepsilon\widehat{\tau}_{3}-i\Delta\widehat{\tau}_{2}}{\sqrt{\Delta^{2}-\varepsilon^{2}}}\sin\varphi_{\mathbf{n}}\right),\label{eq:17}\end{equation}
 where $\varphi_{\mathbf{n}}=k_{\mathrm{F}}\left|\mathbf{n}\right|+\delta$
and the particular forms for the scalar {}``envelop'' function $\mathcal{F}_{0,\mathbf{n}}\left(\varepsilon\right)$
and the phase shift $\delta$ depend on the system dimensionality:\begin{equation}
\mathcal{F}_{0,\mathbf{n}}\left(\varepsilon\right)=\frac{v}{\sqrt{1+v^{2}}}\frac{\sqrt{2}{\textrm{e}}^{-\left|\mathbf{n}\right|/r_{\varepsilon}}}{\sqrt{\pi k_{\mathrm{F}}\left|\mathbf{n}\right|}},\quad\cot\delta=\frac{1-v}{1+v},\label{eq:17a}\end{equation}
 for 2D, and\begin{equation}
\mathcal{F}_{0,\mathbf{n}}\left(\varepsilon\right)=\frac{v}{\sqrt{1+v^{2}}}\frac{{\textrm{e}}^{-\left|\mathbf{n}\right|/r_{\varepsilon}}}{k_{\mathrm{F}}\left|\mathbf{n}\right|},\quad\cot\delta=1/v,\label{eq:17b}\end{equation}
 for 3D, with the energy dependent decay length $r_{\varepsilon}=a^{2}k_{\mathrm{F}}/(\pi\rho_{0}\sqrt{\Delta^{2}-\varepsilon^{2}})$.

The following analysis is essentially simplified, introducing matrices
of the structure:\begin{equation}
\widehat{M}(a,b)=a+b\frac{\varepsilon\widehat{\tau}_{3}-i\Delta\widehat{\tau}_{2}}{\sqrt{\Delta^{2}-\varepsilon^{2}}},\label{eq:18}\end{equation}
 which form a closed algebra with the product rule for the $a,b$
components:\begin{equation}
\widehat{M}\left(a,b\right)\widehat{M}\left(a^{\prime},b^{\prime}\right)=\widehat{M}\left(aa^{\prime}-bb^{\prime},ab^{\prime}+ba^{\prime}\right).\label{eq:19}\end{equation}
 In this notation, the interaction matrices, Eq. \ref{eq:17}, are
presented as\begin{equation}
\widehat{A}_{0,\mathbf{n}}^{0}\approx\mathcal{F}_{0,\mathbf{n}}\widehat{M}\left(\cos\varphi_{\mathbf{n}},\sin\varphi_{\mathbf{n}}\right),\label{eq:19a}\end{equation}
 and then Eq. \ref{eq:19} implies an important formula for their
arbitrary product:\begin{eqnarray}
 &  & \prod_{i=1}^{q}\widehat{A}_{0,\mathbf{n}_{i}}=\nonumber \\
 &  & =\left(\prod_{i=1}^{q}\mathcal{F}_{0,\mathbf{n}_{i}}\right)\widehat{M}\left(\cos\sum_{i=1}^{q}\varphi_{\mathbf{n}_{i}},\sin\sum_{i=1}^{q}\varphi_{\mathbf{n}_{i}}\right).\label{eq:19b}\end{eqnarray}
 This formula permits to reduce an arbitrary polynomial of $\widehat{A}_{0,\mathbf{n}}$
matrices to a single $\widehat{M}$-matrix whose arguments are polynomials
of $\mathcal{F}_{0,\mathbf{n}}$ functions. The next important property
of $\widehat{M}$-matrices is that the determinant:\begin{equation}
\det\left[1-\widehat{M}\left(a,b\right)\right]=\left(1-a\right)^{2}+b^{2}\label{eq:20}\end{equation}
 can be zero only if the components are $a=1$ and $b=0$ simultaneously.

\section{Impurity clusters and in-gap density of states }

The above developed techniques permit to quantify the effects of impurity
clusters in quasiparticle spectrum. Thus, we conclude that the necessary
condition for the pair contribution, Eq. \ref{eq:16}: \begin{eqnarray}
 &  & \det\left[1-\left(\widehat{A}_{0,\mathbf{n}}^{0}\right)^{2}\right]=\nonumber \\
 &  & =\det\left[1-\widehat{M}\left(\mathcal{F}_{0,\mathbf{n}}^{2}\cos2\varphi_{\mathbf{n}},\mathcal{F}_{0,\mathbf{n}}^{2}\sin2\varphi_{\mathbf{n}}\right)\right]=0\label{eq:21}\end{eqnarray}
 is only possible if $\varphi_{\mathbf{n}}=\pi q$, $q=1,\:2,\dots$,
and $\mathcal{F}_{0,\mathbf{n}}^{2}=1$. But, taking into account
the exponential factor ${\textrm{e}}^{-\left|\mathbf{n}\right|/r_{\varepsilon}}<1$
in Eqs. \ref{eq:17a},\ref{eq:17b}, this requires that:\[
\frac{2}{\pi}\frac{v^{2}}{1+v^{2}}>\pi q-\arctan\frac{1+v}{1-v},\]
 for 2D, or \[
\frac{v^{2}}{1+v^{2}}>\pi q-\arctan v,\]
 for 3D, which can not be fulfilled at any $v$ and $q\geq1$. Hence
there is no contribution to DOS within the s-wave gap from impurity
pairs ($\rho_{2}\left(\varepsilon\right)\equiv0$), the same as from
the single-impurity T-matrix, Eq. \ref{eq:15}.

\begin{figure}
\includegraphics[%
  scale=0.7]{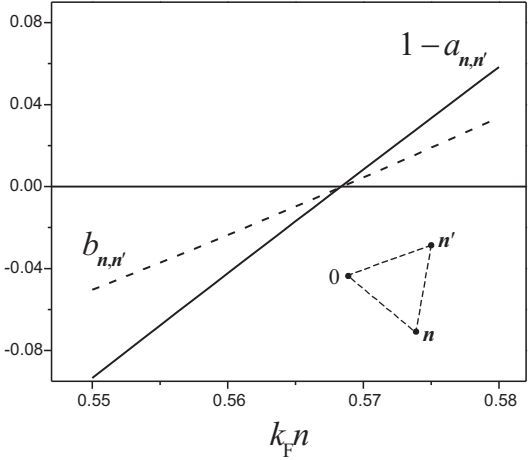}

\caption{\label{cap:1} An example of simultaneous solution of the two conditions,
$a_{\mathbf{n},\mathbf{n}^{\prime}}=1$ and $b_{\mathbf{n},\mathbf{n}^{\prime}}=0$,
necessary for impurity triples to contribute into the in-gap DOS of
a planar SC system. Inset: the chosen geometry of equilateral triangle
$0,\mathbf{n},\mathbf{n}^{\prime}$.}
\end{figure}

But for the next, triple term of GE, which contains the inverse of
the matrix (see, e.g., Ref. \cite{ip}):\begin{eqnarray}
 &  & 1-2\widehat{A}_{0,\mathbf{n}}^{0}\widehat{A}_{0,\mathbf{n}^{\prime}}^{0}\widehat{A}_{\mathbf{n},\mathbf{n}^{\prime}}^{0}-\left(\widehat{A}_{0,\mathbf{n}}^{0}\right)^{2}-\left(\widehat{A}_{0,\mathbf{n}^{\prime}}^{0}\right)^{2}-\left(\widehat{A}_{\mathbf{n},\mathbf{n}^{\prime}}^{0}\right)^{2}=\nonumber \\
 &  & =1-\widehat{M}\left(a_{\mathbf{n},\mathbf{n}^{\prime}},b_{\mathbf{n},\mathbf{n}^{\prime}}\right),\label{eq:22}\end{eqnarray}
 the conditions by Eq. \ref{eq:20} turn already realizable. They
are explicitly formulated as follows:\begin{eqnarray}
 &  & a_{\mathbf{n},\mathbf{n}^{\prime}}=2\mathcal{F}_{0,\mathbf{n}}\mathcal{F}_{0,\mathbf{n}^{\prime}}\mathcal{F}_{\mathbf{n},\mathbf{n}^{\prime}}\cos\left(\varphi_{\mathbf{n}}+\varphi_{\mathbf{n}^{\prime}}+\varphi_{\mathbf{n}-\mathbf{n}^{\prime}}\right)+\nonumber \\
 &  & +\mathcal{F}_{0,\mathbf{n}}^{2}\cos2\varphi_{\mathbf{n}}+\mathcal{F}_{0,\mathbf{n}^{\prime}}^{2}\cos2\varphi_{\mathbf{n}^{\prime}}+\nonumber \\
 &  & +\mathcal{F}_{\mathbf{n},\mathbf{n}^{\prime}}^{2}\cos2\varphi_{\mathbf{n}-\mathbf{n}^{\prime}}=1,\label{eq:23}\end{eqnarray}
 and \begin{eqnarray}
 &  & b_{\mathbf{n},\mathbf{n}^{\prime}}=2\mathcal{F}_{0,\mathbf{n}}\mathcal{F}_{0,\mathbf{n}^{\prime}}\mathcal{F}_{\mathbf{n},\mathbf{n}^{\prime}}\sin\left(\varphi_{\mathbf{n}}+\varphi_{\mathbf{n}^{\prime}}+\varphi_{\mathbf{n}-\mathbf{n}^{\prime}}\right)+\nonumber \\
 &  & +\mathcal{F}_{0,\mathbf{n}}^{2}\sin2\varphi_{\mathbf{n}}+\mathcal{F}_{0,\mathbf{n}^{\prime}}^{2}\sin2\varphi_{\mathbf{n}^{\prime}}+\nonumber \\
 &  & +\mathcal{F}_{\mathbf{n},\mathbf{n}^{\prime}}^{2}\sin2\varphi_{\mathbf{n}-\mathbf{n}^{\prime}}=0.\label{eq:24}\end{eqnarray}
 The easiest localization of quasiparticles of course is expected
at the very edge of the gap: $\varepsilon\rightarrow\Delta$, where
${\textrm{e}}^{-|\mathbf{n}|/r_{\varepsilon}}\rightarrow1$. Then,
e.g., for $v\approx1.728$ it is achieved with $|\mathbf{n}|=|\mathbf{n}^{\prime}|=|\mathbf{n}-\mathbf{n}^{\prime}|\approx0.566k_{\mathrm{F}}^{-1}$,
as shown in Fig. 1, and it can be also reached for close values of
$v$ by small adjustments of the distances.

Once such a pole exists, we can introduce in the configuration space
of triangles $0,\mathbf{n},\mathbf{n}^{\prime}$, in the vicinity
of this pole, the effective variables:\[
r=\sqrt{\left(1-a_{\mathbf{n},\mathbf{n}^{\prime}}\right)^{2}+b_{\mathbf{n},\mathbf{n}^{\prime}}^{2}},\quad\varphi=\arctan\frac{b_{\mathbf{n},\mathbf{n}^{\prime}}}{1-a_{\mathbf{n},\mathbf{n}^{\prime}}},\]
 and a certain $z$, independent of $r,\,\varphi$, arriving at the
general form for $\rho_{3}\left(\varepsilon\right)$ as an integral\begin{eqnarray*}
 &  & \rho_{3}\left(\varepsilon\right)=c^{3}{\textrm{Im}}\int\frac{F\left(r,\varphi,z\right)}{r}dr\: d\varphi\: dz=\\
 &  & =2\pi c^{3}\int F\left(0,\varphi,z\right)d\varphi\: dz.\end{eqnarray*}
 Here the function\[
F\left(r,\varphi,z\right)=2(a_{z,\varphi}\cos\varphi+b_{z,\varphi}\sin\varphi)J(r,\varphi,z)\]
 contains the $\widehat{M}$-algebra coefficients $a_{z,\varphi},\: b_{z,\varphi}$
related to a certain matrix numerator in the triple GE term (see its
explicit form in Ref. \cite{ip}) and the Jacobian $J(r,\varphi,z)$
of transition from $\mathbf{n},\mathbf{n}^{\prime}$ to the effective
variables. Apart from technical details of calculation, this contribution
$\sim c^{3}\rho_{{\textrm{F}}}$ can be routinely obtained.

It should be noted however that the above approach uses the asymptotical
form, Eq. \ref{eq:17}, of interaction functions, so it is only justified
if the resulting poles, like that in Fig. \ref{cap:1}, are related
to long distances $n\gg a$. This is easier achieved if $ak_{{\textrm{F}}}\ll1$
for the host crystal, as is the case for HTSC systems, but hardly
for common s-wave superconductors. In the latter case, the finite
in-gap DOS should result more probably from high enough terms of GE.

Another important aspect is the analysis of convergency of matrix
GE's, Eqs. \ref{eq:11a}, \ref{eq:13}, permitting to distinguish
between band-like and localized spectrum areas in the non-uniform
SC system. This can be realized, using the above estimates for the
non-renormalized functions $\widehat{A}_{0,\mathbf{n}}^{0}$ and $\widehat{T}^{0}$
as approximations for the renormalized ones $\widehat{A}_{0,\mathbf{n}}$
and $\widehat{T}$ in the region of band-like states $\varepsilon>\Delta$.

A similar algebraic techniques can be also developed for the d-wave
SC systems, related in that case to the impurity resonance states.
In this situation, the higher DOS corrections $\rho_{2,\ldots}$ to
non-zero $\rho_{0}+\rho_{1}$ within the gap can be not so pronounced
as for the zero-background case of s-wave system. Nevertheless the
important criteria for GE convergence can be obtained, permitting
a more consistent validation of T-matrix approximation and demarcation
between band-like and localized states, compared to the recently suggested
check based on the phenomenological Ioffe-Regel criterion \cite{lp}.

\end{document}